\documentclass[journal]{IEEEtran}
\usepackage{amsmath,amssymb,amsfonts,amsthm}
\usepackage{algorithmic}
\usepackage{algorithm}
\usepackage{graphicx}
\usepackage{array}
\usepackage{cite}
\usepackage{xcolor}
\usepackage{tikz}
\usetikzlibrary{arrows.meta,positioning,fit,backgrounds,decorations.pathreplacing}
\newcommand{\Nr}{N_{\mathrm{R}}}
\newcommand{\Np}{N_{\text{P}}}
\newcommand{\diag}{\text{diag}}
\newcommand{\bLambda}{\mathbf{\Lambda}}
\newcommand{\bPhi}{\mathbf{\Phi}}
\DeclareMathOperator*{\argmax}{arg\,max}
\newcommand{\SNR}{\text{SNR}}
\newcommand{\figref}[1]{Fig.~\ref{#1}}
\newcommand{\tabref}[1]{Table~\ref{#1}}
\newcommand{\secref}[1]{Section~\ref{#1}}
\newcommand{\algref}[1]{Algorithm~\ref{#1}}
\def\bb0{{\mathbb{0}}}
\def\ba{{\mathbf{a}}}
\def\bb{{\mathbf{b}}}
\def\bc{{\mathbf{c}}}
\def\bg{{\mathbf{g}}}
\def\bh{{\mathbf{h}}}
\def\bm{{\mathbf{m}}}
\def\bs{{\mathbf{s}}}
\def\bu{{\mathbf{u}}}
\def\bv{{\mathbf{v}}}
\def\by{{\mathbf{y}}}
\def\bz{{\mathbf{z}}}
\def\b0{{\mathbf{0}}}
\def\bA{{\mathbf{A}}}
\def\bC{{\mathbf{C}}}
\def\bD{{\mathbf{D}}}
\def\bH{{\mathbf{H}}}
\def\bI{{\mathbf{I}}}
\def\bP{{\mathbf{P}}}
\def\bQ{{\mathbf{Q}}}
\def\bR{{\mathbf{R}}}
\def\bU{{\mathbf{U}}}
\def\bW{{\mathbf{W}}}
\def\bbC{{\mathbb{C}}}
\def\bbE{{\mathbb{E}}}
\def\cN{\mathcal{N}}
\renewcommand{\Np}{N_{\mathrm{p}}}
\renewcommand{\SNR}{\mathsf{SNR}}
\providecommand{\tr}{\mathrm{tr}}
\providecommand{\Var}{\mathrm{Var}}
\providecommand{\range}{\mathrm{range}}
\providecommand{\sinc}{\mathrm{sinc}}
\providecommand{\Trans}{{\scriptscriptstyle \mathrm{T}}}
\providecommand{\jj}{\mathrm{j}}
\providecommand{\Users}{U}
\providecommand{\sumu}{\sum_{u=1}^{\Users}}
\providecommand{\Bmax}{B_{\max}}
\providecommand{\Rfull}{R_{\mathrm{full}}}
\providecommand{\DRtgt}{\Delta R_{\mathrm{target}}}
\providecommand{\Lbar}{\bar{L}}
\providecommand{\sigsq}{\sigma^{2}}
\providecommand{\bINr}{\bI_{\Nr}}
\providecommand{\cNc}{\cN_{\bbC}}
\providecommand{\dspace}{\delta_{\mathrm{a}}}
\providecommand{\reff}[1]{r_{#1}^{\mathrm{eff}}}
\providecommand{\bRang}[1]{\bR_{#1}^{\mathrm{ang}}}
\providecommand{\bDh}[1]{\bD_{#1}^{1/2}}
\providecommand{\bUr}[1]{\bU_{#1}^{(r_{#1})}}
\providecommand{\bUrH}[1]{\bU_{#1}^{(r_{#1})*}}
\providecommand{\bhr}[1]{\tilde{\bh}_{#1}}
\providecommand{\hbR}[1]{\hat{\bR}_{#1}}
\providecommand{\eigd}[1]{\bU_{#1}\bLambda_{#1}\bU_{#1}^*}
\providecommand{\score}[2]{\rho_{#1}(#2)}
\providecommand{\dscore}[2]{\Delta\rho_{#1}(#2)}
\providecommand{\Scap}[2]{S_{#1}(#2)}
\providecommand{\Vcap}[2]{V_{#1}(#2)}
\providecommand{\Tdis}[2]{T_{#1}(#2)}
\providecommand{\bIr}[1]{\bI_{r_{#1}}}

\newtheorem{theorem}{Theorem}
\newtheorem{corollary}{Corollary}
\newtheorem{remark}{Remark}

\begin{document}

\title{Two-Timescale Receive Combining for Fronthaul-Constrained Uplink
XL-MIMO with Visibility Regions}

\author{Jaebum Park,~\IEEEmembership{Graduate Student Member,~IEEE}
  and Robert W. Heath Jr.,~\IEEEmembership{Fellow,~IEEE}%
\thanks{This work has been submitted to the IEEE for possible publication. Copyright may be transferred without notice, after which this version may no longer be accessible.}
\thanks{This material is based upon work supported in part by the National Science Foundation under Grant No. NSF ECCS-2414678 and CCP-2549148.}
\thanks{J. Park and R. W. Heath, Jr. are with the Department of Electrical
and Computer Engineering, University of California at San Diego, La Jolla,
CA 92093, USA (emails: \{jp261, rwheathjr\}@ucsd.edu).}%
}

\maketitle
\bstctlcite{IEEEexample:BSTcontrol}

\begin{abstract}
Extremely large-scale multiple-input multiple-output (XL-MIMO) arrays are hard to deploy in virtualized radio access networks, because the fronthaul data rate grows with the number of antennas when the radio unit forwards every antenna sample. We propose a two-timescale receiver that allocates the per-user combining ranks under the fronthaul stream budget from the channel statistics. The radio unit applies the long-timescale transform, after which the distributed unit applies a minimum mean-squared-error receiver in the reduced dimension. For disjoint visibility regions, we prove that the reduced-dimension receiver decouples across users. We characterize the sum-rate loss to second order and show that the greedy allocation, which we call saturating capture, minimizes its leading term. In simulations, the allocation gains over a uniform-rank baseline, matches a budget-trimmed threshold rule at a small number of users and exceeds it as the array and the user count grow. The gain grows with the heterogeneity of the per-user eigenvalue profiles and vanishes for identical profiles. Under out-of-cell interference offset by at least half a window width, the gain survives with clean covariances, which window-restricted estimation keeps for disjoint windows.
\end{abstract}

\section{Introduction}
\label{sec:intro}

Extremely large-scale multiple-input multiple-output (MIMO), called XL-MIMO, is the evolution of MIMO to the larger arrays required for the upper mid-band between 7 and 24\,GHz~\cite{3gpp38820,bjornson2024}. In a virtualized radio access network, a radio unit (RU) hosts the XL-MIMO array and forwards the received samples over a fronthaul link to a distributed unit (DU), whose baseband processing runs in software on commercial off-the-shelf (COTS) servers~\cite{polese2023,park2026simd}. When the RU forwards one stream per antenna, the fronthaul load grows in proportion to the array size and exceeds the capacity of the link~\cite{3gpp38801,huang2018}. A natural solution is to combine at the RU before forwarding, as the O-RAN 7.2x split places combining in the radio unit~\cite{oranWG4CUS}. The RU does not estimate the channel, so its combiner is sent over the control plane and held for many slots. Such a combiner is built from long-term quantities, the channel statistics in this paper, and must divide a fixed number of forwarded streams among users whose channels occupy different numbers of eigenmodes.

In this paper, we develop a two-stage method for allocating the fronthaul stream budget over a large aperture whose users have different visibility regions~\cite{decarvalho2020}, so that the received energy is spatially non-stationary across the array. The visibility region sets the spatial correlation of a user's channel. The covariance is confined to the antennas in the region, and the number of eigenmodes it occupies grows with the width of the region and with the angular spread. Users therefore differ in power, in angle and in how many eigenmodes their channels occupy, and the allocation exploits that difference. On the long timescale, the DU keeps as many leading eigenvectors of each user's covariance as its allocated rank and stacks them into the combining matrix that the RU applies to every received data vector. A greedy allocation gives each stream to the largest marginal rate gain, the discrete analogue of water-filling over the per-user eigenvalue profiles. On the short timescale, the DU applies a minimum mean-squared-error (MMSE) receiver to the forwarded streams, once per slot. The system is a single radio unit with a uniform linear array that serves single-antenna users on one subcarrier of an orthogonal frequency-division multiplexing (OFDM) uplink, with far-field one-ring correlation inside each visibility window. The decoupling and loss results assume disjoint visibility regions, while the simulation also covers overlapping regions, where we validate the characterization numerically.

\subsection{Prior work}
\label{sec:related}

Prior work used the cluster visibility of large-aperture channel models to analyze linear receivers, to schedule users, and to estimate the region of a user from pilots. In such a model, a user delivers appreciable energy to only part of the array, which is its visibility region~\cite{decarvalho2020}. The term comes from the COST 2100 channel model, where a visibility region is the set of user positions from which a cluster is visible, and for a physically large array the roles swap and the region becomes the set of antennas that a user reaches~\cite{liu2012,flordelis2020}. Our prior work derived signal-to-interference-plus-noise ratio (SINR) expressions for conjugate beamforming and zero forcing under such a channel~\cite{ali2018}, user partitions reduced the cost of partial zero forcing~\cite{zhi2023}, and scheduling users with non-overlapping regions onto shared pilots raised random-access throughput~\cite{marinello2022}. These receivers and schedulers use every antenna. Subarray receivers for extra-large arrays used the non-stationarity to confine detection to the subarrays illuminated by a user and exchanged soft estimates with a central unit~\cite{amiri2018,amiri2022,rodrigues2020}. None of these studies placed the receiver under a stream budget or chose how many dimensions each user keeps. On the acquisition side, the work in~\cite{xu2024,tang2023} recovered the visibility region together with the channel from pilots, the input a region-aware receiver needs, without considering what the receiver should do with it under a fronthaul budget.

One approach to reducing the fronthaul load compresses each forwarded stream, a line the cloud radio access network literature surveys in~\cite{park2014fronthaul,peng2015fronthaul}. Such a design adapts the mantissa length of a block floating-point format per beam to an error-vector-magnitude target~\cite{bychkov2024}. The load then still scales with the number of forwarded streams, because every antenna or beam keeps a stream. Other work optimized the compression jointly with the transmit beamforming under per-link bit budgets~\cite{zhou2016}. Both designs allocated bits over a dimension fixed in advance, without dividing it among the users.

\begin{table}[t]
\caption{Prior work compared with this paper along four design axes (VR: visibility region; C\&F: compress-and-forward)\label{tab:priorwork}}
\begin{center}
    {\renewcommand{\arraystretch}{1.05}%
    \scriptsize\setlength{\tabcolsep}{1.8pt}%
        \begin{tabular}{@{}>{\raggedright\arraybackslash}p{0.44\columnwidth}|>{\centering\arraybackslash}p{0.263\columnwidth}|c|c|c@{}}
        \hline\hline
        Line of work & Granularity & \begin{tabular}[c]{@{}c@{}}Time-\\scale\end{tabular} & \begin{tabular}[c]{@{}c@{}}VR\\model\end{tabular} & \begin{tabular}[c]{@{}c@{}}Budget\\coupling\end{tabular} \\
        \hline
        VR receivers~\cite{decarvalho2020,ali2018,zhi2023,marinello2022,xu2024,tang2023} & full array & single & yes & no\\
        Subarray receivers~\cite{amiri2018,amiri2022,rodrigues2020} & per subarray, fixed & single & yes & no\\
        Quantization~\cite{zhou2016,bychkov2024} & per link, fixed dim. & single & no & no\\
        Dim.\ reduction~\cite{liu2015scf,huang2018,wiffen2022} & per node, fixed or CSI & single & no & no\\
        Per-AP transform~\cite{kim2026} & per AP, fixed & single & no & no\\
        JSDM line~\cite{adhikary2013} & per group, fixed & two & no & no\\
        Hybrid C\&F~\cite{liu2019} & per node, fixed & two & no & no\\
        Statistical selection~\cite{cebeci2025,ouyang2022sel,ouyang2022cap,liu2020beam,marinello2020,asgharimoghaddam2019} & per user, fixed or target & two & partly & no\\
        \textbf{This paper} & \textbf{per user, adaptive} & \textbf{two} & \textbf{yes} & \textbf{yes} \\
        \hline\hline
        \end{tabular}
    }
\end{center}
\end{table}

A second approach to reducing the fronthaul load forwards fewer streams than antennas instead of fewer bits per stream, so that the number of streams kept, the reduced dimension, is the design variable. The line began with spatial compression and forward, in which each remote radio head projected its received vector onto the signal subspace of its instantaneous channel~\cite{liu2015scf}. A truncated conditional Karhunen-Lo\`{e}ve transform at each remote radio head compressed the received vector before quantization, with a decentralized variant in which each radio head computed its filter from its own instantaneous channel and the large-scale coefficients of the network~\cite{wiffen2022}. Moving part of the zero-forcing receiver into the RU under a functional split reduced the number of effective streams that crossed the fronthaul~\cite{huang2018}. Closest to our paper, the work in~\cite{kim2026} optimized the linear transform of each access point (AP) jointly with the fronthaul compression and made the fronthaul overhead of uplink cell-free XL-MIMO independent of the antenna count. That design fixes the output dimension of each AP, computes the transform from the instantaneous channel in every coherence block, and models no visibility region within an AP. The reduction designs chose the reduced dimension per node from the instantaneous channel or set it in advance, over a stationary channel. No prior work on stream reduction allocated a per-user rank under a shared budget or used the visibility regions to choose the kept subspace.

Two-timescale designs split the processing into a stage that uses the channel statistics and a stage that uses the instantaneous channel. Joint spatial division and multiplexing (JSDM) projected each user group onto the dominant eigen-subspace of its channel covariance and served the reduced channel with an instantaneous precoder~\cite{adhikary2013}. On the uplink, a hybrid compress-and-forward design paired a statistical stage with an instantaneous digital stage at the remote radio head and water-filled quantization bits over a fixed reduced dimension~\cite{liu2019}. Channel aging degrades any design tied to instantaneous channel knowledge~\cite{truong2013}, whereas the covariance drifts on a longer timescale, which one study tracked through its temporal correlation~\cite{daei2024}. The two-timescale designs set the reduced dimension in advance, per group by a threshold or per radio head by a rule, and assume stationary statistics.

Statistical selection methods choose the kept subspace from the long-term statistics, either as a set of beams or as a set of antennas. These methods are our baselines. Processing each user in a fixed window of discrete Fourier transform (DFT) beams concentrated the interference in a few eigenmodes and supported near-optimal reduced-dimension MMSE detection~\cite{cebeci2025}. Antenna selection kept a hard subset chosen from the long-term statistics instead of a subspace~\cite{ouyang2022sel}, and the sum-rate scaling of such a subset was characterized in~\cite{ouyang2022cap}. Under a visibility-region channel, selecting the antennas with the most received energy from the long-term fading raised the energy efficiency of XL-MIMO~\cite{marinello2020}. In~\cite{liu2020beam}, each user was served on beams chosen from the angular support and the large-scale gains. A two-stage uplink receiver chose the dimension of each user's outer beamformer from deterministic equivalents of the MMSE projection onto sectors of DFT beams, so that the dimension followed the angular overlap among the users~\cite{asgharimoghaddam2019}. That rule set a per-user complexity target with no budget shared across the users. The selection methods fix the per-user dimension in advance or set it per user without a shared total, whereas under a visibility-region channel the effective rank of a user grows with its window width and angular spread, so the users offer unequal numbers of useful eigenmodes. A fixed dimension assigns streams to users whose capture has saturated while other users still hold strong eigenmodes. No prior method allocates the per-user rank under a shared stream budget from the visibility-restricted statistics or characterizes the resulting sum-rate loss. \tabref{tab:priorwork} compares the prior work with our paper along the compression granularity, the timescale, the visibility-region model, and the coupling of the budget across users.

\subsection{Contributions}

The contributions of this paper are as follows.
\begin{itemize}
\item \textbf{Statistics-only allocation algorithm.} We develop saturating-capture rank allocation, a greedy allocation over the per-user eigenvalue profiles that is exactly optimal for the separable single-stream surrogate of the sum rate. Our algorithm assigns each stream to the user whose next eigenmode adds the most single-stream rate. It uses only the long-term statistics and eigendecompositions that other covariance-based methods also compute, such as JSDM~\cite{adhikary2013} and fixed beamspace processing~\cite{cebeci2025}. The allocation itself adds a negligible cost on top of them, one comparison over the users per forwarded stream.
\item \textbf{Sum-rate loss and out-of-cell interference.} We prove that the reduced-dimension MMSE receiver decouples across users with disjoint visibility regions. From the decoupling, we derive a second-order characterization of the sum-rate loss imposed by the fronthaul stream budget, show that the leading term is a separable concave function of the per-user ranks, so that the saturating-capture allocation minimizes it exactly, and correct the bound by a leakage term based on region overlap. The decoupling, and with it the characterization, survives out-of-cell interference that stays outside the visibility windows. For the pilot contamination that a sequence-sharing interferer adds to the covariance estimate, we apply the orthogonal-support property of~\cite{bjornson2018} to the estimate itself. Restricting the estimate to the visibility window removes the contamination exactly when the two windows are disjoint, while a subtraction with the interferer covariance known removes it when they overlap.
\item \textbf{Evaluation of the gain and its limits.} We compare the allocation with uniform-rank, eigenvalue-threshold, fixed-window beamspace, and antenna-selection baselines on a visibility-region channel. The allocation removes a large part of the rate gap left by the uniform-rank baseline at the budgets of interest. Against a budget-trimmed eigenvalue-threshold rule, it matches the rate at eight users and pulls ahead as the system grows to 1024 antennas and 64 users. The advantage over that rule is therefore mainly an optimality guarantee for the leading loss term. We find that the gain concentrates where the users are heterogeneous in visibility. As predicted by our analysis, the gains disappear in the homogeneous limit. Though we assume perfect covariances in most of the paper, we find that the gains hold even when the statistics are estimated from a limited snapshot window. Under out-of-cell interference, the gain survives where the interferer windows are offset from the user windows by at least half a width. Window-restricted estimation recovers most of the pilot-contamination loss at a half-width offset.
\end{itemize}

We extend a conference version~\cite{park2026asilomar}, which introduced the combiner and evaluated it on one geometry. We add the decoupling theorem, the second-order loss characterization with its optimality and accuracy properties, the leakage-corrected bound for overlapping regions, the estimated-statistics and multi-cell studies, and the placement of both stages on the O-RAN split.

\emph{Notation:} Bold lower- and uppercase letters denote vectors and matrices, $(\cdot)^{\Trans}$ the transpose, $(\cdot)^*$ the Hermitian transpose, $\diag(\cdot)$ a diagonal matrix, $\tr(\cdot)$ the trace, $\|\cdot\|$ the norm, $\cNc(\bm,\bC)$ the circularly symmetric complex Gaussian distribution with mean $\bm$ and covariance $\bC$, and $\bbE[\cdot]$ expectation.

\section{System model}
\label{sec:model}

\begin{figure}[t]
\centering
\includegraphics[width=0.8\columnwidth]{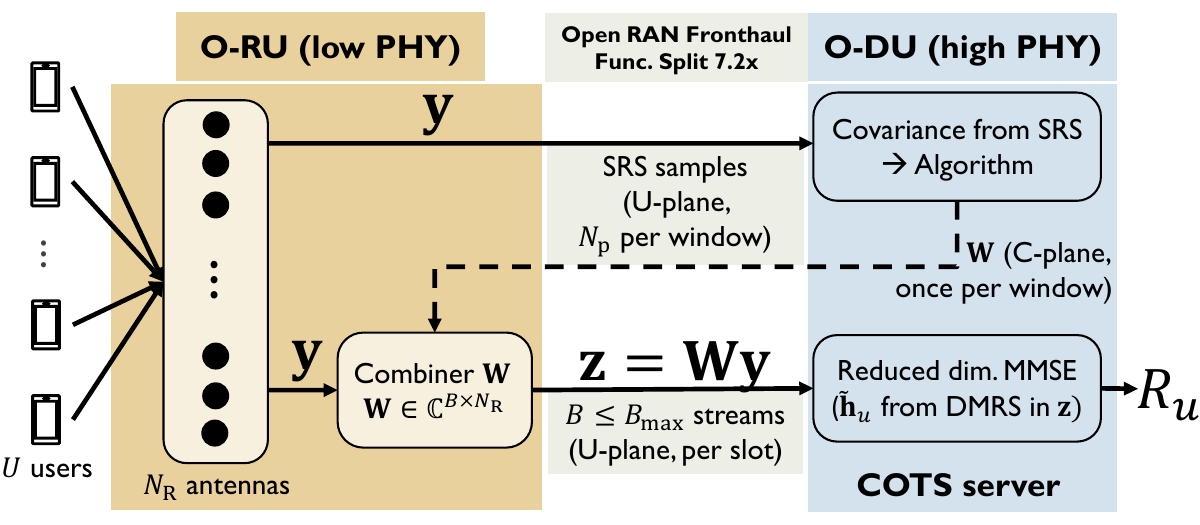}
\caption{Uplink XL-MIMO receive architecture on the O-RAN 7.2x split. The O-RU applies the long-timescale combining matrix $\bW$ and forwards $B$ streams in place of $\Nr$ antenna samples. The O-DU estimates the covariances $\{\bR_u\}$ from the $\Np$ sounding snapshots forwarded in each statistics window, computes $\bW$, sends it to the O-RU over the control plane once per window, and runs the reduced-dimension MMSE receiver once per slot.}
\label{fig:system}
\end{figure}

\figref{fig:system} depicts the uplink of a single XL-MIMO RU whose array serves several single-antenna users and forwards $B$ streams to the DU over a capacity-limited fronthaul, where $B$, the forwarded stream count, is smaller than the number of antennas. The RU applies a combining matrix to every received vector, while the DU applies an MMSE receiver to every forwarded vector. Only their update rates differ. The combining matrix follows from the per-user covariances and is recomputed once per statistics window of many coherence blocks, while the MMSE receiver follows from the instantaneous reduced-dimension channel and is recomputed once per slot. \secref{sec:acq} describes how the DU estimates the covariances and sends the combining matrix to the RU. A functional split is the point in the baseband chain at which the RU hands over to the DU. Among the splits that 3GPP cataloged~\cite{3gpp38801}, the O-RAN Alliance standardized the lower-layer 7.2x split, which places precoding and combining in the RU, called the O-RAN radio unit (O-RU), and the equalization in the DU, called the O-RAN distributed unit (O-DU), so that the fronthaul transports spatial streams rather than antenna samples~\cite{oranWG4CUS}. The forwarded stream count is then the natural budget unit.

\subsection{Signal model}
We consider an array of $\Nr$ antennas that serves $\Users$ single-antenna users on the uplink of an OFDM system. We write the model for one subcarrier, indexed by $k$, and then say why the rest of the paper drops the index. Let $s_u[k]\sim\cNc(0,1)$ denote the data symbol of user $u$ on subcarrier $k$, let $p_u>0$ denote its transmit power, and let $\bh_u[k]\in\bbC^{\Nr}$ denote its channel on that subcarrier. Let $\bv[k]\sim\cNc(\b0,\sigsq\bINr)$ denote the noise that the array adds, with noise power $\sigsq$ per antenna. We collect the channels in $\bH[k]=[\bh_1[k],\dots,\bh_{\Users}[k]]$, the powers in $\bP=\diag(p_1,\dots,p_{\Users})$, and the symbols in $\bs[k]=[s_1[k],\dots,s_{\Users}[k]]^{\Trans}$. The received vector $\by[k]\in\bbC^{\Nr}$ on subcarrier $k$ in one channel use is then
\begin{align}
\by[k] = \sumu\sqrt{p_u}\,\bh_u[k] s_u[k] + \bv[k] = \bH[k]\bP^{1/2}\bs[k] + \bv[k].
\label{eq:rx}
\end{align}
We assume that the spatial correlation of a user does not change across the band, so that the covariance $\bbE[\bh_u[k]\bh_u[k]^*]$ is the same on every subcarrier. The combining matrix of \secref{sec:proposed} is built from these covariances alone, so one matrix serves every subcarrier, while the DU computes its MMSE receiver per subcarrier from $\bH[k]$, as in any OFDM receiver. We therefore drop the index $k$ and write the model for one subcarrier in the rest of the paper.

\begin{figure}[t]
\centering
\begin{minipage}[t]{0.49\columnwidth}\centering\includegraphics[width=\linewidth]{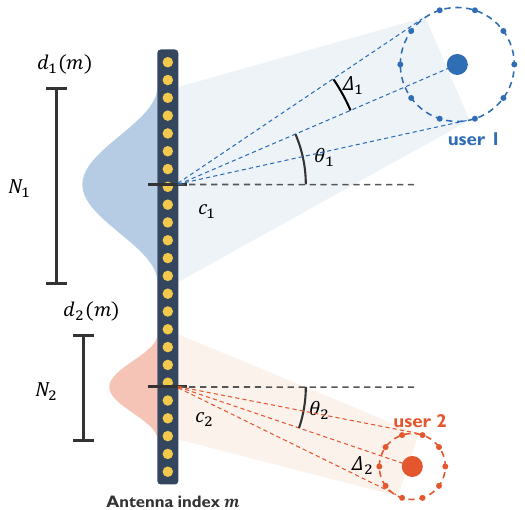}\\[-2pt]{\footnotesize (a)}\end{minipage}\hfill
\begin{minipage}[t]{0.49\columnwidth}\centering\includegraphics[width=\linewidth]{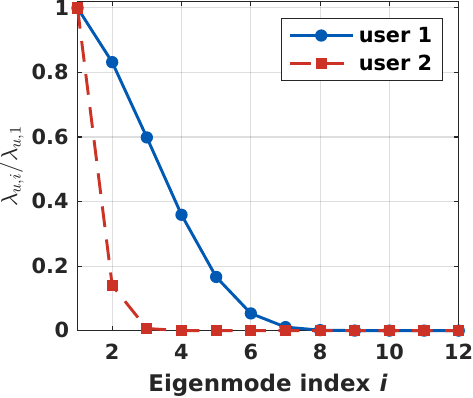}\\[-2pt]{\footnotesize (b)}\end{minipage}
\caption{Visibility-region geometry and the per-user eigenvalue profiles it produces. (a)~Each user illuminates a window of $N_u$ antennas centered at antenna $c_u$ with the raised-cosine taper $d_u(m)$ of~\eqref{eq:taper}. (b)~Normalized eigenvalues $\lambda_{u,i}/\lambda_{u,1}$ of the covariance~\eqref{eq:cov} under the geometry of (a), which decay within a few modes and differ across users.}
\label{fig:geometry}
\end{figure}

\subsection{Channel model}
We assume that the aperture is large enough so that each user is visible over a limited window of antennas instead of over the whole array, as shown in \figref{fig:geometry}(a). This is the extra-large regime of~\cite{decarvalho2020}, in which users differ in the part of the array they reach. An array that every user reaches through different clusters per segment corresponds instead to full overlap, which \secref{sec:results} evaluates as full overlap. We also assume that the scatterers of a user belong to a single cluster, so that the antennas over which the user is visible form one contiguous window. This assumption keeps the covariance of a user a single tapered block, so that the per-user eigenvalue profile of \figref{fig:geometry}(b) and the rank allocation of \secref{sec:proposed} stay tractable.

We build the covariance of a user from two deterministic factors, an energy taper across the array and an angular correlation among the antennas inside the window. Let $d_u(m)$ denote the average energy that user $u$ impinges on antenna $m$, let $N_u$ denote the number of antennas in its window, and let $c_u$ denote the index of the antenna at the window center. We take the taper in the graded raised-cosine form
\begin{align}
d_u(m)=
\begin{cases}
\tfrac{1}{2}\left(1+\cos\tfrac{2\pi(m-c_u)}{N_u}\right), & |m-c_u|\le\tfrac{N_u}{2},\\
0, & \text{otherwise},
\end{cases}
\label{eq:taper}
\end{align}
so the received energy rolls off smoothly toward the window edges. In this model, the visibility region of user $u$ is the set of antennas with $d_u(m)>0$, namely $\{m:|m-c_u|<N_u/2\}$, since the taper vanishes at $|m-c_u|=N_u/2$. We say that two visibility regions are disjoint when they share no antenna. We use a graded taper rather than an on-off mask because measurements along large arrays show the received power of a user varying gradually over the array~\cite{payami2012,gao2015}. In those measurements, the variation came from shadowing that differed across the array, while the waves inside a window of ten antennas were treated as planar. Our taper carries the first feature, and the in-window correlation of the next paragraph carries the second. The COST 2100 channel model, developed within the European Cooperation in Science and Technology (COST) Action 2100, gives each visibility region a transition region in which the cluster gain rises smoothly from zero~\cite{liu2012,flordelis2020}. \secref{sec:results} repeats the evaluation under the hard mask to show that the taper shape does not drive the gain. The diagonal matrix $\bD_u=\diag(d_u(1),\dots,d_u(\Nr))$ collects the taper.

An alternative model keeps the energy flat inside the window and cuts it at the edge~\cite{decarvalho2020}, the hard visibility mask $\bar d_u(m)=1$ for $|m-c_u|\le N_u/2$ and $0$ otherwise. Windows with a flat middle and cosine edges interpolate between them.

We then model the correlation inside the window. We take the antennas to form a uniform linear array whose spacing in wavelengths is $\dspace=1/2$. Because scatterers around user $u$ sit at different angles, the array receives the user over a spread of angles of arrival. Let $\theta_u$ denote the mean angle of arrival of user $u$ and let $\Delta_u$ denote its one-sided angular spread. Let $\bRang{u}$ denote the angular correlation matrix, whose entries the one-ring model gives as~\cite{shiu2000,adhikary2013}
\begin{align}
[\bRang{u}]_{mn} =\frac{1}{2\Delta_u}\int_{-\Delta_u}^{\Delta_u} e^{\,\jj2\pi\dspace(m-n)\sin(\theta_u+\varphi)}\,d\varphi .
\label{eq:onering}
\end{align}
Let $\sinc(x)=\sin(\pi x)/(\pi x)$. For a small spread, the standard first-order approximation is $[\bRang{u}]_{mn}\approx e^{\,\jj\pi(m-n)\sin\theta_u}\, \sinc\left((m-n)\Delta_u\cos\theta_u\right)$.

Finally, we combine the visibility region and the antenna correlation into the covariance of user $u$ as
\begin{align}
\bR_u = \bDh{u}\bRang{u}\bDh{u},
\label{eq:cov}
\end{align}
which generalizes the binary visibility-region covariance to a graded energy profile~\cite{ali2018}. The symmetric scaling by $\bDh{u}$ keeps $\bR_u$ Hermitian and positive semidefinite. Its diagonal satisfies $[\bR_u]_{mm}=d_u(m)$, so that $\bD_u$ alone determines how the average energy is distributed across the array, while $\bRang{u}$ determines the correlation among the in-window entries. For two users with disjoint visibility regions, $\bD_j\bD_u=\b0$, and therefore $\bR_j\bR_u=\b0$.

A user seen through several separated clusters has one window per cluster and a covariance that is the sum of the corresponding blocks. Because the blocks occupy disjoint antennas, the sum keeps their eigenvectors together with the union of their eigenvalue profiles. The allocation of \secref{sec:proposed} applies without change, and Theorem~\ref{thm:decouple} holds when no window of one user shares an antenna with a window of another user. We do not evaluate users with several clusters.

Users of neighboring cells also reach the array. Let $U_I$ denote their number, let $\bh_e^{o}$ denote the channel of interferer $e$, let $\bR_e^{o}=\bbE[\bh_e^{o}\bh_e^{o*}]$ denote its covariance, built as in~\eqref{eq:cov} from its own window and angles, and let $q_e$ denote its received power. We collect the interference in $\bQ=\sum_{e=1}^{U_I}q_e\bR_e^{o}$. The single-cell evaluation sets $U_I=0$, and \secref{sec:multicell} adds one interferer per user.

Let $\bg_u\sim\cNc(\b0,\bINr)$ collect the independent unit-variance small-scale fading coefficients of user $u$, redrawn in every coherence block. The channel of user $u$ is then the correlated Rayleigh-fading vector with $\bbE[\bh_u\bh_u^*]=\bR_u$,
\begin{align}
\bh_u = \bR_u^{1/2}\bg_u .
\label{eq:chan}
\end{align}
Let $\lambda_{u,i}$ denote the $i$th largest eigenvalue of $\bR_u$ and $\bu_{u,i}$ the corresponding eigenvector. We write the eigendecomposition as $\bR_u=\eigd{u}$ with $\bU_u=[\bu_{u,1},\dots,\bu_{u,\Nr}]$ and $\bLambda_u=\diag(\lambda_{u,1},\dots,\lambda_{u,\Nr})$.

The per-user eigenvalues decay within a few modes at a rate that depends on the visibility geometry. Let $\reff{u}$ denote the effective rank of $\bR_u$, the number of dominant eigenvalues that together capture 95\% of the trace $\tr(\bR_u)$. For a stationary linear aperture of $N_u$ half-wavelength elements and angular support $2\Delta_u$, the covariance concentrates its energy on roughly $\reff{u}\approx N_u\Delta_u\cos\theta_u$ dominant eigenmodes, the spatial degrees-of-freedom result of~\cite{poon2005} that JSDM also used~\cite{adhikary2013}. The in-window angular correlation determines the decay rate, whereas the taper shape changes it little.

Restricting a user to its window makes $\reff{u}$ small, because $N_u$ is a fraction of $\Nr$, and unequal across users, because they differ in window width, angular spread, and angle. In \figref{fig:geometry}(b), the wide-window user, with $N_u=48$, $\Delta_u=6.9^\circ$, and $\theta_u=10^\circ$, offers roughly six dominant modes, while the narrow-window user, with $N_u=18$, $\Delta_u=4.6^\circ$, and $\theta_u=35^\circ$, offers barely more than one. The effective rank is a property of the channel and counts the eigenmodes the channel offers, whereas the rank $r_u$ is a design variable and counts the eigenmodes the RU keeps under the fronthaul budget.

The received power of a user is confined to its visibility region. At antenna $m$, the average received power of user $u$ over the small-scale fading is $p_u\bbE|[\bh_u]_m|^2=p_u[\bR_u]_{mm}=p_ud_u(m)$, which vanishes outside the window. We define the per-antenna received signal-to-noise ratio (SNR) of user $u$ as the array average
\begin{align}
\SNR_u=\frac{p_u\tr(\bR_u)}{\Nr\sigsq}.
\label{eq:snrdef}
\end{align}
In the evaluation, we fix $\SNR_u$ and solve~\eqref{eq:snrdef} for the power, $p_u=\SNR_u\,\Nr\sigsq/\tr(\bR_u)$, so that a narrow visibility region raises $p_u$ rather than reducing the received SNR. This calibration holds the received SNR fixed as the visibility geometry varies, so the geometry changes the shape of the eigenvalue profile without changing the received power.

\subsection{Two-timescale reduction and performance metric}

On the long timescale, the DU assigns user $u$ a rank $r_u\in\{0,\dots,\mathrm{rank}(\bR_u)\}$ and keeps the truncated eigen-subspace $\bUr{u}=[\bu_{u,1},\dots,\bu_{u,r_u}]$ spanned by the $r_u$ dominant eigenvectors of $\bR_u$. The forwarded stream count is then $B=\sumu r_u$. The combining matrix stacks the kept subspaces as
\begin{align}
\bW=\left[\bUr{1},\dots,\bUr{\Users}\right]^* \in\bbC^{B\times \Nr}.
\label{eq:reduce}
\end{align}
On the short timescale, the RU forwards the $B$ complex samples $\bz=\bW\by$ per channel use to the DU, where a per-user MMSE receiver recovers the symbols. Let $\bhr{u}=\bW\bh_u$ denote the reduced-space effective channel of user $u$. The DU estimates it per slot from the demodulation reference signal (DMRS) inside $\bz$, since $\bW$ acts on the antennas and leaves the pilot positions unchanged. We apply $\bW$ to~\eqref{eq:rx} and group the result, from the viewpoint of user $u$, into the desired signal, the interference, and the projected noise,
\begin{align}
\bz=\sqrt{p_u}\,\bhr{u} s_u +\sum_{j\ne u}\sqrt{p_j}\,\bhr{j} s_j+\bW\bv.
\label{eq:zsplit}
\end{align}
Let $\bPhi=\bbE[\bz\bz^*] =\sum_{j=1}^{\Users}p_j\bhr{j}\bhr{j}^* +\sigsq\bW\bW^*$ denote the covariance of the forwarded vector at a fixed channel realization, whose interference-plus-noise part for user $u$, after removing that user's own term, is $\bPhi-p_u\bhr{u}\bhr{u}^*$. Under out-of-cell interference, the interferer channels add their terms $q_e\bW\bh_e^{o}\bh_e^{o*}\bW^*$ to $\bPhi$, which the DU estimates from the forwarded samples without knowing those channels. We write the rate of user $u$ under the MMSE receiver as
\begin{align}
R_u(\bW)=\log_2\!\left(1+p_u\,\bhr{u}^* \left(\bPhi-p_u\bhr{u}\bhr{u}^*\right)^{-1} \bhr{u}\right).
\label{eq:rate}
\end{align}
We measure performance by the uplink sum rate $\sum_u R_u(\bW)$ in expectation over the small-scale fading, with the full-dimension receiver $\bW=\bINr$ allowed by an unconstrained fronthaul as the reference throughout. Let $\Rfull=\sum_u \bbE\left[R_u(\bINr)\right]$ denote its sum rate.

\subsection{Covariance acquisition}
\label{sec:acq}
We assume that the covariances $\{\bR_u\}$ are known at the DU on the long timescale and fixed over a statistics window, while the instantaneous channels $\{\bh_u\}$ change every coherence block. This separation is reasonable because the covariances follow the user positions and the scattering geometry, which change orders of magnitude more slowly than the fading. 

The DU estimates the covariances from pilots on the long timescale. Let $\hat{\bh}_u[n]$ denote the pilot estimate of $\bh_u$ in the $n$th sounded coherence block, and let $\Np$ denote the number of such snapshots in one statistics window, which observe independent small-scale fading realizations while $\bR_u$ stays fixed. Averaging their outer products gives the sample covariance
\begin{align}
\bR_u \approx \hbR{u}=\frac{1}{\Np}\sum_{n=1}^{\Np} \hat{\bh}_u[n]\,\hat{\bh}_u[n]^*.
\label{eq:scov}
\end{align}
Prior work on joint visibility-region detection and channel estimation confined the estimate to the detected window, which may improve it further~\cite{xu2024,tang2023}. \secref{sec:results} quantifies the loss down to $\Np=25$ snapshots. Let $\bar{\bD}_u=\diag(\bar d_u(1),\dots,\bar d_u(\Nr))$ denote the hard window mask of user $u$ from \secref{sec:model}. \secref{sec:ooc} uses it to remove the pilot contamination that a neighboring cell adds to the estimate.

In 3GPP New Radio (NR) terms, a sounded coherence block is a sounding reference signal (SRS) occasion and the short-timescale processing is the per-slot equalization~\cite{3gppTS38.211}. The DU estimates the covariances from the SRS samples the RU forwards over the user plane. The RU receives the combining matrix over the control plane once per statistics window.

\subsection{Problem statement}
\label{sec:problem}
We make four further modeling assumptions to facilitate the analysis. First, we assume a single RU with a far-field one-ring in-window correlation. At a carrier of 7\,GHz, the widest window of 44 half-wavelength elements is $0.94$\,m long and has a Rayleigh distance of about $41$\,m. Users beyond that distance are in the far field of each window, while much of the cell is in the near field of the full $2.7$\,m aperture. Second, we measure the fronthaul cost by the stream count $B=\sum_u r_u$ and treat the per-stream quantizer resolution as a separate variable, which \secref{sec:checks} revisits under a fixed total bit budget. Third, we let the users differ in power and in visibility geometry, as they do under a large aperture. Fourth, each user has one antenna. For a user with several antennas and $L_u$ layers under a Kronecker channel, a transmit precoder scales the receive covariance and keeps its eigenvectors, which leaves the allocation unchanged apart from the constraint $r_u\ge L_u$, under which the $L_u$ layers share the kept block. The loss analysis of that case remains open. The design uses only the covariances $\{\bR_u\}$, the powers $\{p_u\}$, and the noise power $\sigsq$.

As the fronthaul supports at most $\Bmax$ streams while the array offers $\Users\Nr$ candidate eigenmodes across the users, we choose how many of them to keep for each user. The rank $r_u$ counts the received dimensions forwarded over the fronthaul to recover the single stream of user $u$. A single-antenna user carries one layer, so the MMSE receiver at the DU recovers one symbol per user. The rank $r_u$ still exceeds one because $\bW$ follows from the statistics rather than from the instantaneous channel. The RU therefore forwards the eigenmodes of $\bR_u$ that together capture most of the channel energy.

Given the statistics $\{\bR_u\}$, the powers $\{p_u\}$, the noise power $\sigsq$, and a fronthaul stream budget $\Bmax$, we formulate the rank-allocation problem as
\begin{align}
&\max_{\{r_u\}}\ \sumu\bbE\left[R_u(\bW)\right]\nonumber\\
&\ \text{s.t.}\ \ \sumu r_u\le \Bmax,\ \ r_u\in\{0,\dots,\mathrm{rank}(\bR_u)\},
\label{eq:alloc}
\end{align}
where the expectation is over the small-scale fading $\{\bg_u\}$, $\bW$ follows from $\{r_u\}$ as in~\eqref{eq:reduce}, and $R_u(\bW)$ is the rate in~\eqref{eq:rate}. The variables are the integer per-user ranks, since we fix the short-timescale receiver to the MMSE receiver of~\eqref{eq:rate}.

We define the cost of the fronthaul constraint as the gap to the full-dimension reference,
\begin{align}
\Delta R(\{r_u\}) = \Rfull - \sumu\bbE\left[R_u(\bW)\right].
\label{eq:loss}
\end{align}
Because both terms of~\eqref{eq:loss} average over the same fading, the gap is the rate lost to the stream budget alone. The loss is nonnegative, because MMSE on $\bW\by$ cannot exceed MMSE on $\by$, is invariant to an invertible transform of the subspace, and vanishes when the row space of $\bW$ contains every $\bh_u$.

The loss depends on the per-user ranks $\{r_u\}$ through a simple trade, in which raising $r_u$ by one uses one fronthaul stream and keeps one more eigenmode of user $u$, of energy $\lambda_{u,r_u+1}$. Because the energy that a stream recovers depends on the user and on the rank already assigned, the budget should go to the users whose eigenvalue tails still hold energy, which a uniform split ignores. \secref{sec:proposed} develops a greedy allocation and \secref{sec:analysis} characterizes $\Delta R$ to second order.

\section{Rank allocation by saturating capture}
\label{sec:proposed}

We propose a greedy fronthaul stream allocation, which gives each stream to the user whose next eigenmode adds the most single-stream rate. Because a user transmits one data stream, the rate that its next eigenmode adds depends on the energy that the kept subspace already holds for that user. Once the subspace holds most of a user's statistical energy, one more eigenmode adds almost no rate, while the first eigenmode of a user with no streams yet adds its full single-stream rate. The allocation therefore stops feeding a saturated user and moves the remaining streams to the users whose next eigenmode still adds rate, without any fairness constraint. We call the resulting greedy allocation \textit{saturating capture}. In this section, we state the surrogate problem that it solves exactly, describe its implementation, and give its complexity.

\subsection{A separable problem and greedy rank allocation}
Let $\Scap{u}{r}=\sum_{i\le r}\lambda_{u,i}$ denote the statistical energy of user $u$ captured by its first $r$ eigenmodes. We define the single-stream rate score of user $u$ at rank $r$ as
\begin{align}
\score{u}{r}=\log_2\!\left(1+\frac{p_u\,\Scap{u}{r}}{\sigsq}\right),
\label{eq:proxy}
\end{align}
which treats the captured energy as the signal power of one stream and ignores multi-user interference. The MMSE stage of~\eqref{eq:rate} suppresses the interference on the short timescale. Let $\eta_u=p_u/\sigsq$ denote the transmit power of user $u$ normalized by the noise power, so that $\score{u}{r}=\log_2(1+\eta_u\Scap{u}{r})$. Summing the per-user scores and replacing the objective of~\eqref{eq:alloc} by that sum gives the separable problem
\begin{align}
&\max_{\{r_u\}}\ \sumu\score{u}{r_u}\nonumber\\
&\ \text{s.t.}\ \ \sumu r_u\le \Bmax,\ \ r_u\in\{0,\dots,\mathrm{rank}(\bR_u)\}.
\label{eq:surrogate}
\end{align}
We define the marginal gain of one more stream for user $u$ as
\begin{align}
\dscore{u}{r}=\score{u}{r+1}-\score{u}{r} =\log_2\!\left(1+\frac{p_u\lambda_{u,r+1}}{\sigsq+p_u \Scap{u}{r}}\right).
\label{eq:marginal}
\end{align}
Because the eigenvalues decay, $\dscore{u}{r}$ is decreasing in $r$. Problem~\eqref{eq:surrogate} is then a separable concave allocation with integer variables, for which the greedy that always assigns the next stream to the largest marginal gain is exactly optimal by marginal analysis~\cite{fox1966}. Corollary~\ref{cor:opt} restates this result for the score. With a separable objective and decreasing marginals, the greedy selects the $\Bmax$ largest marginal gains. The combining matrix $\bW$ stacks the selected eigenvectors as in~\eqref{eq:reduce}.

Ranking the candidate eigenmodes by the marginal gain in~\eqref{eq:marginal} is not the same as ranking them by energy. Any increasing function of the per-eigenmode energy $p_u\lambda_{u,i}$ alone, the logarithm included, orders the candidates as the energy does and returns the energy allocation, whereas the marginal gain also depends on the accumulated capture $\Scap{u}{r}$ and therefore changes with the rank that the user has already received. Energy ranking is the low-SNR linearization of the score, valid only when $\eta_u\Scap{u}{r}\ll1$. For a user with $p_u\Scap{u}{r}\gg\sigsq$ the marginal gain decreases as $p_u\lambda_{u,r+1}/\left(p_u\Scap{u}{r}+\sigsq\right)$, whereas the marginal gain of a user with $r_u=0$ equals its full single-stream rate $\log_2(1+p_u\lambda_{u,1}/\sigsq)$. 

A gain over the uniform split requires eigenvalue profiles that differ across users, which the visibility-region channel produces through the per-user window widths, angular spreads, and angles. In the homogeneous limit, where all users share one eigenvalue profile and one power, the greedy returns exactly the uniform split.

\subsection{Implementation and complexity analysis}
\algref{alg:satgreedy} uses only the long-timescale quantities, since its inputs are the covariances and the powers rather than the instantaneous channel. The DU runs the algorithm and sends the combining matrix $\bW$ over the control plane to the RU, which applies $\bW$ to each received data vector. The ranks $\{r_u\}$ stay at the DU, which uses them to divide the $B$ forwarded streams among the users. In the algorithm, $m_u$ holds the current marginal gain $\dscore{u}{r_u}$ of user $u$, which line~9 recomputes after each assignment.

\begin{algorithm}[t]
\caption{Saturating-capture rank allocation}
\label{alg:satgreedy}
\begin{algorithmic}[1]
\STATE \textbf{Input:} covariances $\{\bR_u\}$, window masks $\{\bar{\bD}_u\}$, powers $\{p_u\}$, noise power $\sigsq$, stream budget $\Bmax$ \STATE \textbf{Output:} ranks $\{r_u\}$, combining matrix $\bW$ \STATE Restrict $\bR_u\leftarrow\bar{\bD}_u\bR_u\bar{\bD}_u$ for all $u$ \STATE Compute $\bR_u=\eigd{u}$ for all $u$ \STATE Set $r_u = 0$, $S_u = 0$, and $m_u = \dscore{u}{0} = \log_2(1+p_u\lambda_{u,1}/\sigsq)$ for all $u$ \WHILE{$\sum_u r_u<\Bmax$} \STATE $u^\star = \argmax_u\, m_u$ \STATE Set $r_{u^\star} = r_{u^\star}+1$ and $S_{u^\star} = S_{u^\star}+\lambda_{u^\star,r_{u^\star}}$ \STATE Update $m_{u^\star} = \Delta\rho_{u^\star}(r_{u^\star})$ from~\eqref{eq:marginal} \ENDWHILE \STATE Form $\bW=[\bUr{1},\dots,\bUr{\Users}]^*$ as in~\eqref{eq:reduce} \RETURN $\{r_u\}$, $\bW$
\end{algorithmic}
\end{algorithm}

Each statistics window needs one eigendecomposition per user at the DU, a cost of $O(\Users\Nr^3)$ per window that every covariance-based baseline incurs. Because $\bR_u$ vanishes outside its window, the decomposition can act on the $N_u\times N_u$ block, at a cost of $O(\sum_u N_u^3)$. The allocation itself costs $O(\Bmax\log \Users)$ with a max-heap over the $\Users$ marginals, after which forming $\bW$ copies $\Bmax$ eigenvectors. On the short timescale, the receiver is the reduced-dimension MMSE of~\eqref{eq:rate} on $B$ streams rather than $\Nr$, which lowers the per-block computation at the DU. The allocation therefore adds almost nothing to the computation of a covariance-based receiver.
\section{Sum-rate loss analysis}
\label{sec:analysis}

In this section, we characterize the loss $\Delta R(\{r_u\})$ and show that its leading term is built from the single-stream rate score $\score{u}{r}$ of \secref{sec:proposed}. We start from an exact decoupling result for disjoint visibility regions, expand it into a deterministic second-order characterization, and use it to prove \algref{alg:satgreedy} optimal for the leading term. We close the section with a budget rule, a bound for overlapping visibility regions, and the effect of out-of-cell interference. Throughout, we write $\Scap{u}{r}=\sum_{i\le r}\lambda_{u,i}$ for the captured statistical energy and $\Vcap{u}{r}=\sum_{i\le r}\lambda_{u,i}^2$ for the sum of squared eigenvalues.

\subsection{Decoupling and second-order characterization of the loss}

Disjoint visibility regions decouple the reduced-dimension MMSE rates exactly, so that the loss splits into per-user scalar terms. Of interest, the decoupling holds for every channel realization and not only on average.

\begin{theorem}[Exact decoupling for disjoint visibility regions]\label{thm:decouple}
If the visibility regions are disjoint, $\bR_j\bR_u=\b0$ for $j\ne u$, and $r_u\le\mathrm{rank}(\bR_u)$ for every $u$, then for every channel realization the reduced-dimension MMSE rates decouple, $R_u(\bW)=\log_2(1+\eta_u\|\bUrH{u} \bh_u\|^2)$. Let $g_{u,i}=[\bU_u^*\bg_u]_i\sim\cNc(0,1)$ denote the fading coefficient of eigenmode $i$. Let $\Sigma_u=\sum_{i\le r_u}\lambda_{u,i}|g_{u,i}|^2$ denote the captured instantaneous energy and $\Sigma_u^{\perp}=\sum_{i>r_u}\lambda_{u,i}|g_{u,i}|^2$ the discarded instantaneous energy. The loss in~\eqref{eq:loss} is then
\begin{align}
\Delta R=\sumu\bbE\!\left[\log_2\!\left( 1+\frac{\eta_u \Sigma_u^{\perp}}{1+\eta_u\Sigma_u}\right)\right].
\label{eq:exactloss}
\end{align}
\end{theorem}
\begin{IEEEproof}
See Appendix~\ref{app:decouple}.
\end{IEEEproof}

The loss in~\eqref{eq:exactloss} separates into $\Users$ per-user scalar terms with no cross-user quantity left, each of which depends on the ratio of that user's discarded instantaneous energy to its captured instantaneous energy. That expression still involves an expectation over the captures, which we remove by expanding the logarithm to second order about the mean capture.

\begin{theorem}[Second-order characterization]\label{thm:char}
Under the assumptions of Theorem~\ref{thm:decouple}, let $\Lbar=\sum_u[\score{u}{\Nr}-\score{u}{r_u}]$ be the mean-capture loss, the loss built from the score~\eqref{eq:proxy} with each capture replaced by its mean, and let the residual $\varepsilon$ denote the sum over the users of the difference between the third-order Lagrange remainders of the per-user expansions at full rank and at rank $r_u$, defined in the proof. Then
\begin{align}
\Delta R={}&\Lbar+\frac{\log_2e}{2}\sumu\eta_u^2\!\left[ \tfrac{\Vcap{u}{r_u}}{(1+\eta_u\Scap{u}{r_u})^2}-\tfrac{\Vcap{u}{\Nr}}{(1+\eta_u\Scap{u}{\Nr})^2}\right]\nonumber\\
&+\varepsilon.
\label{eq:char}
\end{align}
\end{theorem}
\begin{IEEEproof}
Let $\Sigma_u(r)=\sum_{i\le r}\lambda_{u,i}|g_{u,i}|^2$ denote the instantaneous capture of user $u$ at rank $r$, so that $\Sigma_u(r_u)=\Sigma_u$ and $\Sigma_u(\Nr)=\Sigma_u+\Sigma_u^{\perp}$. The squared modulus of a $\cNc(0,1)$ variable is exponential with unit mean and unit variance, so $\Sigma_u(r)$ is a weighted sum of independent unit-mean exponential variables with mean $\bbE[\Sigma_u(r)]=\Scap{u}{r}$ and variance $\Var(\Sigma_u(r))=\Vcap{u}{r}$. Let $f(x)=\log_2(1+\eta_ux)$, which we expand about the mean before taking the expectation. The first-order term vanishes because $\bbE[\Sigma_u(r)-\Scap{u}{r}]=0$. By Taylor's theorem with the Lagrange remainder,
\begin{align}
\bbE[f(\Sigma_u(r))] =f\left(\Scap{u}{r}\right)+\tfrac12 f''\left(\Scap{u}{r}\right)\Vcap{u}{r} +\varepsilon_u(r),
\end{align}
where $\varepsilon_u(r)=\tfrac16\,\bbE[f'''(\xi_u)(\Sigma_u(r)-\Scap{u}{r})^3]$ for some $\xi_u$ between $\Scap{u}{r}$ and $\Sigma_u(r)$. The remainder $\varepsilon_u(r)$ is finite because $f'''$ is bounded. With $f(\Scap{u}{r})=\score{u}{r}$ and $f''(x)=-\eta_u^2\log_2e/(1+\eta_ux)^2$, the expansion becomes
\begin{align}
\bbE\log_2\left(1+\eta_u\Sigma_u(r)\right) =\score{u}{r}-\frac{\eta_u^2\Vcap{u}{r}\log_2e}{2(1+\eta_u\Scap{u}{r})^2}+\varepsilon_u(r).
\label{eq:expcap}
\end{align}
Subtracting~\eqref{eq:expcap} at $r=r_u$ from~\eqref{eq:expcap} at $r=\Nr$ and summing over the users yields~\eqref{eq:char} with $\varepsilon=\sum_u[\varepsilon_u(\Nr)-\varepsilon_u(r_u)]$.
\end{IEEEproof}

Replacing $f'''(\xi_u)$ by $f'''(\Scap{u}{r})$ in the remainder of the proof gives the third-order term $\tfrac23\log_2e\sum_{i\le r}\lambda_{u,i}^3/(1/\eta_u+\Scap{u}{r})^3$ at each rank. The residual $\varepsilon$ therefore scales as the difference of this term at $r=\Nr$ and $r=r_u$ summed over the users, a scaling we verify numerically. Therefore, the characterization is accurate when each capture sums many eigenmodes, since the sum then stays close to its mean. It degrades at ranks of one or two, where a single eigenmode determines the capture and fluctuates as much as its mean.

\begin{corollary}[Optimality of \algref{alg:satgreedy}]\label{cor:opt}
\algref{alg:satgreedy} minimizes $\Lbar$ over all integer allocations $\{r_u\}$ with $\sum_u r_u\le \Bmax$.
\end{corollary}
\begin{IEEEproof}
Because the term $\sum_u\score{u}{\Nr}$ in $\Lbar$ does not depend on the ranks, minimizing $\Lbar$ under $\sum_u r_u\le\Bmax$ is the surrogate problem~\eqref{eq:surrogate}. Since $\score{u}{0}=0$, the objective telescopes into a sum of increments, $\sum_u\score{u}{r_u}=\sum_u\sum_{i=0}^{r_u-1}\dscore{u}{i}$, each nonnegative and nonincreasing in $i$ by~\eqref{eq:marginal}. An allocation therefore selects the first $r_u$ increments of each user, and the sum of $\Bmax$ such increments is largest when they are the $\Bmax$ largest increments over all users, which form a feasible allocation because the increments of each user are nonincreasing. The greedy takes the largest remaining increment at each step and selects exactly these increments, which is the marginal-analysis argument of~\cite{fox1966}.
\end{IEEEproof}

The optimality of Corollary~\ref{cor:opt} applies to the leading term $\Lbar$ and not to the exact loss in~\eqref{eq:exactloss}. The second-order term in~\eqref{eq:char} measures the difference between the two, which is largest at the small per-user ranks where the budget binds.

\subsection{Budget selection and overlapping visibility regions}
\label{sec:designrule}
The characterization also yields a budget rule for the number of fronthaul streams required for a target sum-rate loss $\DRtgt$. The rule evaluates $\Lbar$ at the greedy ranks for increasing $\Bmax$ and selects the smallest budget at which $\Lbar$ does not exceed $\DRtgt$. It needs only the covariances.

The direction of the error of the rule depends on the geometry. Because the rule tests the target against $\Lbar$ instead of $\Delta R$, it selects a budget at least as large as the exact loss requires where $\Lbar$ exceeds $\Delta R$ and may fall short where $\Lbar$ underestimates $\Delta R$. Under disjoint visibility regions, the second-order term in~\eqref{eq:char} is nonnegative whenever $\eta_u\Scap{u}{r_u}\ge1$ for every user, since $\eta_u^2\Vcap{u}{r}/(1+\eta_u\Scap{u}{r})^2$ is then nonincreasing in $r$ for $r\ge r_u$. In that regime $\Lbar$ underestimates the loss up to the residual $\varepsilon$. For a user with $r_u=0$, or at low SNR, the second-order term is negative and $\Lbar$ overestimates the loss. The condition $\eta_u\Scap{u}{r_u}\ge1$ holds for every user and budget of \secref{sec:results}. Under overlapping regions, the decoupling is also inexact, which may instead raise $\Lbar$ above the exact loss, as \secref{sec:results} finds on the heterogeneous geometry.

As Theorems~\ref{thm:decouple} and~\ref{thm:char} assume disjoint visibility regions, we bound the loss for overlapping regions by adding an interference term to a bound for the disjoint case. Overlapping regions make the decoupling of Theorem~\ref{thm:decouple} inexact and load the forwarded streams with cross-user interference that the MMSE receiver has to suppress. Rows of $\bW$ outside the kept eigenmodes could aid that suppression. Under a fixed $\Bmax$, however, such rows come out of the user ranks, which the score of~\eqref{eq:proxy} does not value. We do not pursue them here.

Let $\Tdis{u}{r}=\Scap{u}{\Nr}-\Scap{u}{r}$ denote the discarded statistical energy of user $u$ at rank $r$. Under the hypotheses of Theorem~\ref{thm:decouple}, the per-user term of~\eqref{eq:exactloss} satisfies
\begin{align}
\bbE\left[\log_2\left(1+\frac{\eta_u\Sigma_u^{\perp}}{1+\eta_u\Sigma_u}\right)\right] &\le \bbE\left[\log_2(1+\eta_u\Sigma_u^{\perp})\right],
\label{eq:dropcap}\\
\bbE\left[\log_2(1+\eta_u\Sigma_u^{\perp})\right] &\le \log_2\left(1+\eta_u\Tdis{u}{r_u}\right),
\label{eq:jensen}
\end{align}
where~\eqref{eq:dropcap} drops $\Sigma_u\ge0$ from the denominator and~\eqref{eq:jensen} applies Jensen's inequality to the concave logarithm, since the discarded instantaneous energy $\Sigma_u^{\perp}$ has mean $\Tdis{u}{r_u}$. Summing~\eqref{eq:jensen} over the users, we obtain the disjoint bound as
\begin{align}
\Delta R\le\sumu\log_2\!\left(1+\eta_u\Tdis{u}{r_u}\right).
\label{eq:disjointbound}
\end{align}
Let $\ell_u=\sum_{j\ne u}p_j\tr\left(\bUrH{u}\bR_j \bUr{u}\right)$ denote the statistical power of the other users in the subspace that the RU keeps for user $u$. For disjoint visibility regions, $\bR_j\bUr{u}=\b0$ for $j\ne u$ and $\ell_u=0$. As the visibility regions overlap, the kept subspaces of different users align and $\ell_u$ grows.

\begin{remark}[Leakage-corrected bound for overlapping visibility regions]\label{rem:leak}
For overlapping visibility regions, we correct the bound~\eqref{eq:disjointbound} by the leakage $\ell_u$ as
\begin{align}
\Delta R\lesssim\sumu\log_2\!\left(1+\eta_u\Tdis{u}{r_u}\right) +\sumu\log_2\!\left(1+\frac{\ell_u}{\sigsq}\right).
\label{eq:leak}
\end{align}
The second sum vanishes for disjoint visibility regions, where~\eqref{eq:leak} reduces to~\eqref{eq:disjointbound}.
\end{remark}

The corrected bound in~\eqref{eq:leak} is an estimate rather than a theorem, which \secref{sec:results} validates numerically for overlapping regions. Both bounds are loose, because dropping $\Sigma_u$ from the denominator discards the captured energy before Jensen's inequality. The leakage term also counts interference that the MMSE receiver partly removes. We therefore predict the loss with the characterization~\eqref{eq:char}.
\subsection{Out-of-cell interference}
\label{sec:ooc}
Out-of-cell interference enters the receiver at both timescales, adding $\bW\bQ\bW^*$ to the noise covariance of the forwarded vector on the short timescale and contaminating the covariance estimate on the long timescale when a neighboring cell reuses the sounding sequence.

\begin{corollary}[Decoupling under interference outside the windows]\label{cor:ooc}
If $\bQ$ has no support inside any user window, so that $\bar{\bD}_u\bQ=\b0$ for every $u$, then $\bW\bQ\bW^*=\b0$ and Theorems~\ref{thm:decouple} and~\ref{thm:char} hold unchanged.
\end{corollary}
Every row of $\bW$ lies in the column space of one $\bR_u$, which the window of that user supports, so the interference term vanishes from every block and the proofs apply as written. Interference that reaches a window stays in the forwarded dimensions, where a user kept at rank one cannot null it, a case that \secref{sec:multicell} observes.

When a neighboring cell reuses the sounding sequence of user $u$, the sample covariance~\eqref{eq:scov} converges to $\bR_u+(q_{\pi(u)}/p_u)\bR^{o}_{\pi(u)}$, where $\pi(u)$ denotes the interferer that shares the sequence. Two users whose covariances have orthogonal supports can share a sequence without contamination at any array size~\cite{bjornson2018}, which the visibility-region channel realizes whenever their windows are disjoint~\cite{marinello2022}. The window mask applies that property to the estimate. If the taper of interferer $\pi(u)$ vanishes inside the window of user $u$, then $\bar{\bD}_u\hat{\bh}_u[n]$ holds no interferer energy in any snapshot, so the window-restricted estimate $\bar{\bD}_u\hbR{u}\bar{\bD}_u$ equals the contamination-free sample covariance restricted to the window for every $\Np$. Line~3 of \algref{alg:satgreedy} applies the restriction. Where the two windows overlap, the restriction leaves the term $(q_{\pi(u)}/p_u)\bar{\bD}_u\bR^{o}_{\pi(u)}\bar{\bD}_u$ inside the window. When the DU knows that covariance, the knowledge that multi-cell MMSE combining assumes, subtracting the term restores the clean estimate at any overlap. \secref{sec:multicell} quantifies both remedies.

\section{Simulation results}
\label{sec:results}

\begin{figure}[t]
\centering
\includegraphics[width=\columnwidth]{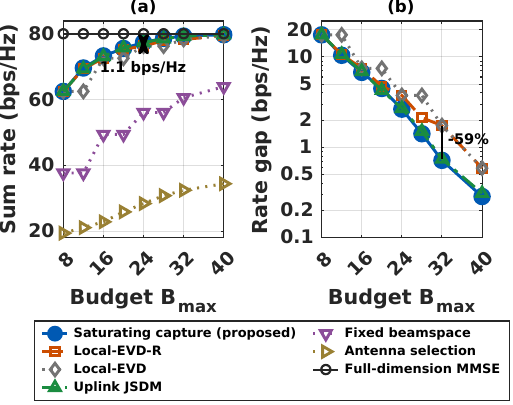}
\caption{(a)~Uplink sum rate versus the fronthaul stream budget $\Bmax$ at SNR 10\,dB for $\Nr=128$ and $\Users=8$, with visibility widths from 17 to 44 antennas, angular spreads from $2.9^\circ$ to $27.2^\circ$, and overlap $\omega=0.5$. The allocation gains $0.72$, $1.11$, and $1.02$\,bps/Hz over Local-EVD-R at $\Bmax=16$, $24$, and $32$, where its ranks at $\Bmax=24$ are $[1\,2\,2\,3\,4\,4\,4\,4]$ against three per user for the uniform split. (b)~Gap to the full-dimension reference for the four statistics-based methods, which the allocation shrinks from $3.8$ to $2.6$\,bps/Hz at $\Bmax=24$ and from $1.7$ to $0.7$\,bps/Hz at $\Bmax=32$.}
\label{fig:budget}
\end{figure}

We evaluate our proposed rank allocation algorithm on a uniform linear array of $\Nr=128$ antennas serving $\Users=8$ users at nominal SNRs of 5 and 10\,dB. Let the nominal SNR denote the value of $\SNR_u$ in~\eqref{eq:snrdef} before the per-user offsets. Quoted numbers refer to 10\,dB unless stated otherwise. We take $\Bmax\ge \Users$ throughout, since a smaller budget cannot serve every user. The per-user widths, SNR offsets, angles, and angular spreads sit on equally spaced grids over their ranges, user $u$ taking the $(u-1)/(\Users-1)$ point of each. Across the users, the widths run from 17 to 44 antennas, the SNR offsets from $-7.5$ to $7.5$\,dB around the nominal value, the mean angles from $-60^\circ$ to $60^\circ$, and the angular spreads from $2.9^\circ$ to $27.2^\circ$. Let $\bar N$ denote the mean window width in antennas across the users, and let $\omega\in[0,1]$ denote the overlap parameter that sets the spacing of the visibility-region centers as $c_{u+1}-c_u=(1-\omega)\bar N$, with the centers placed symmetrically about the middle of the array. At $\omega=0$ neighboring regions meet edge to edge on average, while at $\omega=1$ their centers coincide. Unless stated otherwise, $\omega=0.5$. Each point is an average of $10^3$ fading realizations, shared across the methods. The standard error of the paired difference between two methods is below $0.06$\,bps/Hz in \figref{fig:budget} to \figref{fig:overlap}, $0.1$\,bps/Hz in \tabref{tab:scale}, and $0.12$\,bps/Hz in \figref{fig:multicell}. We treat smaller differences as unresolved.

All reference methods use the same long-term statistics. We do not adapt the transforms of~\cite{kim2026} and~\cite{wiffen2022}, which are computed from the instantaneous channel and cannot be sent over the control plane once per statistics window. The uniform baseline, Local-EVD-R, assigns every user the rank $\lfloor \Bmax/\Users\rfloor$ from the local eigenvalue decomposition (EVD) of its covariance and gives the remaining streams, one each, to the users with the largest next eigenvalue. Plain Local-EVD leaves those streams unused and coincides with it where $\Users$ divides $\Bmax$.

The remaining baselines choose the kept dimensions by an energy threshold, a window of DFT beams, or a subset of antennas. The uplink JSDM baseline adapts the threshold-based eigenspace selection of JSDM~\cite{adhikary2013} to the uplink budget. It assigns each user the smallest rank that captures 90\% of its energy and, when the total exceeds the budget, trims the largest ranks one stream at a time until the budget is met. When the total falls short of the budget, the remaining streams stay unused, as a threshold rule has no further candidates. The fixed beamspace baseline, after~\cite{cebeci2025}, gives each user a contiguous window of $\lfloor \Bmax/\Users\rfloor$ DFT beams centered on its strongest beam. Antenna selection, after~\cite{ouyang2022sel}, forwards the $\Bmax$ antennas with the largest long-term energy $\sum_u p_u[\bR_u]_{mm}$. The full-dimension MMSE receiver is the upper reference.

\subsection{Sum rate versus the stream budget}
The allocation gains about $1$\,bps/Hz over the uniform baseline across the budgets of interest, as \figref{fig:budget}(a) shows for the uplink sum rate at SNR 10\,dB and \figref{fig:budget}(b) for the gap to the full-dimension reference on a logarithmic axis. The ordering of the methods is unchanged at 5\,dB. All eigenspace-based methods reach within 1\,bps/Hz of the full-dimension reference by $\Bmax=40$, one third of the antennas, which confirms the low-rank premise of the channel. Below that budget, the allocation removes up to 59\% of the gap left by the uniform baseline. It holds the low-rank users at one eigenmode and moves the freed streams to the users whose eigenvalue tails are still strong, as the ranks in the caption show.

\subsection{Heterogeneity, overlap, and estimated statistics}
\begin{figure}[t]
\centering
\includegraphics[width=0.8\columnwidth]{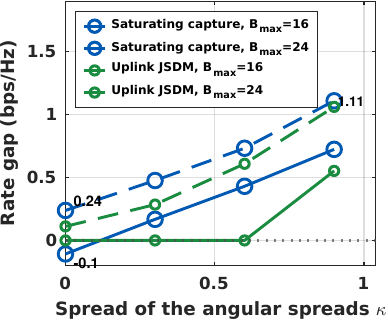}
\caption{Sum-rate gap over Local-EVD-R versus the spread $\kappa$ of the per-user angular spreads at $\Bmax\in\{16,24\}$ and SNR 10\,dB. At $\Bmax=24$ the gap grows from $0.24$\,bps/Hz at zero spread, where the users still differ in visibility width and power, to $1.11$\,bps/Hz at $\kappa=0.9$. At $\Bmax=16$ and zero spread the allocation falls $0.1$\,bps/Hz below the baseline, a deficit above the paired standard error of $0.04$\,bps/Hz.}
\label{fig:heterogeneity}
\end{figure}

The gain grows with the heterogeneity of the per-user eigenvalue profiles and disappears without it, as the sweep in \figref{fig:heterogeneity} shows. Let $\kappa\in[0,1]$ scale the spread of the per-user angular spreads, $\Delta_u=\bar\Delta(1+\kappa\nu_u)$ with $\bar\Delta=15^\circ$ and $\nu_u$ equally spaced on $[-0.9,0.9]$, so that $\kappa=0$ gives every user the same spread and $\kappa=0.9$ is the geometry of \figref{fig:budget}. Sweeping $\kappa$ from 0 to 0.9 widens the per-user effective ranks from the range 3--7 to the range 1--9, and the gap over the uniform baseline grows monotonically with it. With fully homogeneous users, the allocation reduces to the uniform split, since identical eigenvalue profiles give identical marginal gains. At zero spread, the users still differ in visibility width and power, which leaves the small gain at $\Bmax=24$ and the small deficit at $\Bmax=16$ that the caption quantifies.

\begin{figure}[t]
\centering
\includegraphics[width=0.8\columnwidth]{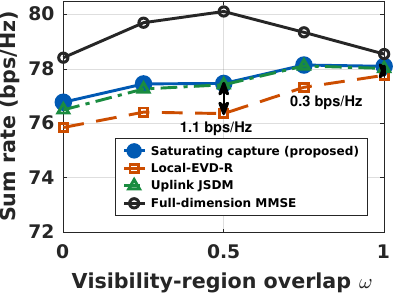}
\caption{Uplink sum rate versus the visibility-region overlap $\omega$ at $\Bmax=24$ and SNR 10\,dB. The gain over Local-EVD-R narrows to $0.3$\,bps/Hz as the overlap grows, where the allocation and uplink JSDM come within $0.1$\,bps/Hz at $\omega=1$.}
\label{fig:overlap}
\end{figure}

The gain persists as the visibility regions move from separated to coinciding, as \figref{fig:overlap} shows, where the allocation does not fall below the uniform baseline at any $\omega$. As $\omega$ grows, the multi-user interference omitted by the capture score also grows, and as the kept subspaces of the users align, the MMSE receiver uses part of its $B$ dimensions to suppress that interference, which narrows the gain.

Replacing the true covariances with the sample estimates~\eqref{eq:scov} costs little rate. We build the allocation from an $\Np$-snapshot sample covariance of noiseless channel draws, so that the finite-sample error is the only loss, and average over 100 snapshot draws. The estimated allocation loses $0.51$, $0.31$, and $0.17$\,bps/Hz against its true-covariance version at $\Np=25$, 50, and 100 snapshots, and clears the uniform baseline built from the same estimates by more than $1$\,bps/Hz, since the allocation needs only the ordering of the accumulated eigenvalue sums, which a few tens of snapshots resolve.

\subsection{Multi-cell interference and pilot contamination}
\label{sec:multicell}
\begin{figure}[t]
\centering
\includegraphics[width=0.8\columnwidth]{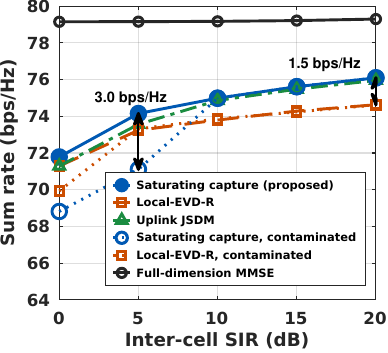}
\caption{Uplink sum rate versus the inter-cell SIR at $\Bmax=24$ and SNR 10\,dB, with one out-of-cell interferer per user that reuses its sounding sequence and a window offset from the user's by half a width ($\omega_I=0.5$). Solid curves use the limiting contaminated covariance ($\Np\to\infty$) restricted to the user windows and dotted curves the same covariance unrestricted. With clean covariances the allocation leads Local-EVD-R by $0.55$ to $1.6$\,bps/Hz at every SIR. The unrestricted covariance reverses the ordering at $5$\,dB and below, $41$ against $46$\,bps/Hz at $-5$\,dB, and the restricted form recovers $65$\,bps/Hz there and $74.2$ against $71.1$ at $5$\,dB. The $-5$\,dB points lie outside the axis.}
\label{fig:multicell}
\end{figure}

The gain survives out-of-cell interference where the interferer windows are offset from the user windows by at least half a width and the covariances stay clean, and window-restricted estimation keeps them clean where the windows are disjoint. We add one interferer per user, the user of a neighboring cell that reuses its sounding sequence, with the width of that user, a window offset from the user's by $(1-\omega_I)N_u$ with $\omega_I\in[0,1]$ the interferer-to-user overlap, an angle and a spread drawn as for the users, and a received SNR lower than the user's by the inter-cell SIR. \figref{fig:multicell} plots the sum rate versus the SIR at $\Bmax=24$ and $\omega_I=0.5$ for the limiting contaminated covariance and its window-restricted form. With clean covariances, the allocation keeps its gain over Local-EVD-R at every SIR and every overlap for the interferer draw of the figure, since the short-timescale MMSE receiver suppresses the interference in the forwarded dimensions. At $\omega_I=1$ the gain depends on the draw, and over ten draws of the interferer angles it turns negative in about half of them at SIRs of $5$\,dB or less, when the interferer of the user kept at rank one falls inside that eigenmode, as \secref{sec:ooc} notes, while at $\omega_I=0.5$ it stays positive in every draw. The contaminated estimate reverses the ordering at $\omega_I=0.5$ and an SIR of $5$\,dB or less, and at $-5$\,dB every method collapses, because the greedy assigns streams to the eigenmodes of the interferer. Window-restricted estimation restores the clean combining matrix exactly at $\omega_I=0$, where the sequence-sharing interferer lies outside the window of its user, and recovers most of the loss at $\omega_I=0.5$, as the caption quantifies. Across the ten draws the window-restricted allocation keeps its lead at $5$\,dB in every draw and at $-5$\,dB in seven of ten, where the part of the interferer inside the window still misleads the greedy, and with $\Np=25$ snapshots the restriction recovers two thirds of that loss at $-5$\,dB. The windows at $\omega_I=1$ do not separate the interferer from the user, so every statistics-based method loses at an SIR of $0$\,dB or less, the allocation most, while from $5$\,dB upward the allocation keeps its lead. The remedy at full overlap is the covariance-aware decontamination of~\cite{yin2013}, the subtraction of \secref{sec:ooc} with the interferer covariance known, which restores the clean covariance at every overlap.

\subsection{Analysis validation}
\begin{figure}[t]
\centering
\includegraphics[width=0.8\columnwidth]{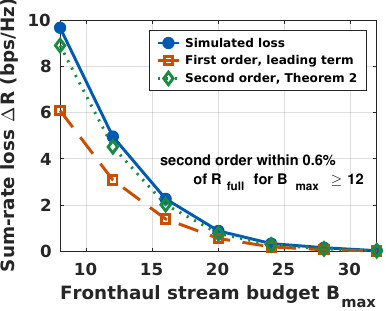}
\caption{Sum-rate loss versus $\Bmax$ with the first- and second-order characterizations of Theorem~\ref{thm:char}, for $\Nr=128$, $\Users=8$, SNR 10\,dB, windows of 14 antennas, and $\omega=0$. For $\Bmax\ge 12$ the second-order characterization predicts the loss to within 0.6\% of the full-dimension rate.}
\label{fig:loss}
\end{figure}

The second-order characterization predicts the simulated loss to within $0.6\%$ of the full-dimension rate for budgets of at least 12 streams, as \figref{fig:loss} shows against the first- and second-order characterizations of Theorem~\ref{thm:char} on a geometry with disjoint visibility regions. The decoupled rate expression of Theorem~\ref{thm:decouple} matches the simulated per-user rates. The first-order term underestimates the loss at tight budgets, and the underestimate shrinks as the budget grows, since each capture then sums more eigenmodes and stays closer to its mean.

The budget rule of \secref{sec:designrule} selects $\Bmax=24$ streams for $\DRtgt=3$\,bps/Hz and $\Bmax=32$ streams for $\DRtgt=1.4$\,bps/Hz on the heterogeneous geometry, from the covariances alone, and the simulated gaps at those budgets in \figref{fig:budget} meet both targets. Because the leading term exceeds the simulated loss at both budgets, an error of the rule costs extra streams rather than a missed target. The heterogeneous geometry at $\omega=0.5$ lies outside the disjoint hypothesis of Theorems~\ref{thm:decouple} and~\ref{thm:char}, where $\Lbar$ is an empirical approximation that this agreement supports. Over $\omega$ from 0 to 1 at $\Bmax=24$, neither the disjoint bound~\eqref{eq:disjointbound} nor the leakage-corrected bound~\eqref{eq:leak} is violated, though both stay one to two orders of magnitude above the simulated loss.

\subsection{Design checks and scaling}
\label{sec:checks}
Six checks show that the comparison is insensitive to the capture-score approximation, the taper shape, the threshold of the JSDM baseline, the nominal SNR, and the distance of the greedy from the exhaustive maximizer, and that streams are the right budget unit. Replacing the capture score $\score{u}{r}$ in~\eqref{eq:proxy} with the exact expected single-stream rate $\bbE[\log_2(1+\eta_u\Sigma_u)]$ changes the sum rate by at most $0.25$\,bps/Hz across the sizes of \tabref{tab:scale}.

Under the hard visibility mask of \secref{sec:model}, the gain over the uniform baseline at $\Bmax=24$ grows from $1.1$ to $1.4$\,bps/Hz, because the mask keeps more energy at the window edges. The taper is therefore not the source of the gain. The 80\%, 90\%, and 95\% energy thresholds of the JSDM baseline give the same sum rate to within $0.01$\,bps/Hz. Over nominal SNRs from $-5$ to 20\,dB the gain over the uniform baseline stays between $1.1$ and $1.2$\,bps/Hz. At $\Nr=32$ and $\Users=4$, enumerating every allocation with $\sum_u r_u=\Bmax$ for $\Bmax\in\{8,12\}$, for disjoint regions and for the default overlap, puts the greedy within $0.03$\,bps/Hz of the exhaustive maximizer. With a uniform quantizer of $b$ bits per complex sample on each forwarded stream, $b/2$ per real component with the step set to clip at three standard deviations of the long-term stream power, and a fixed bit budget $C=Bb$ of $192$ bits per channel use, the sum rate peaks at $b=12$ and $B=16$ and falls at $B=32$ with $b=6$, so streams are the binding unit at $12$ bits or more. Under the additive-noise model, the DU adds the quantization noise to the interference-plus-noise covariance of the MMSE receiver.

\tabref{tab:scale} reports the sum rates and the measured allocation time across five system sizes at $\Bmax=3\Users$ and SNR 10\,dB. Its geometry sets the mean visibility width to $2\Nr/(\Users+1)$, the widest at which the windows tile the array at $\omega=0.5$, so its $(128,8)$ row differs from the geometry of \figref{fig:budget}. The Time column is the long-timescale stage, the per-user eigendecompositions plus the greedy, in MATLAB R2024b on an Intel Core i9-14900KS processor with default threading, and stays within a statistics window of hundreds of milliseconds to seconds, over which the large-scale parameters of the 3GPP channel model stay correlated at pedestrian to vehicular speeds~\cite{3gppTR38.901}. As the users grow with the array, the gain over JSDM rises from twice the paired standard error at $(128,8)$ to seventeen times it at $(1024,64)$.

\begin{table}[t]
\caption{Sum rate (bps/Hz) and long-timescale allocation time\label{tab:scale}}
\begin{center}
    {\renewcommand{\arraystretch}{1.05}%
    \footnotesize\setlength{\tabcolsep}{2.5pt}%
        \begin{tabular}{c|c|c|c|c|c}
        \hline\hline
        $(\Nr,\Users)$ & \begin{tabular}[c]{@{}c@{}}Saturating\\capture\end{tabular} & \begin{tabular}[c]{@{}c@{}}Local-\\EVD-R\end{tabular} & \begin{tabular}[c]{@{}c@{}}Uplink\\JSDM\end{tabular} & \begin{tabular}[c]{@{}c@{}}Full-dimension\\MMSE\end{tabular} & Time (s) \\
        \hline
        $(64,8)$    & 70.7  & 70.5  & 70.4  & 70.8  & 0.001 \\
        $(128,8)$   & 77.8  & 76.6  & 77.7  & 80.0  & 0.003 \\
        $(256,16)$  & 171.6 & 169.2 & 171.3 & 176.2 & 0.028 \\
        $(512,32)$  & 374.4 & 370.2 & 373.7 & 384.2 & 0.28 \\
        $(1024,64)$ & 811.8 & 804.0 & 810.5 & 833.1 & 3.1 \\
        \hline\hline
        \end{tabular}
    }
\end{center}
\end{table}

\subsection{Discussion}
\label{sec:discussion}
Heterogeneity of the per-user eigenvalue profiles is the source of the gain of rank adaptation, as \figref{fig:heterogeneity} and the hard-mask check of \secref{sec:checks} show. Random geometries produce smaller gains than the deterministic geometry of \figref{fig:budget}, which places one user at each end of every range. Over 100 geometries with the widths, angles, spreads, and offsets uniform over those ranges, the gain over Local-EVD-R at $\Bmax=24$ averages $0.38$\,bps/Hz and exceeds twice its paired standard error in 86\% of the draws, while the gain over JSDM averages $0.08$\,bps/Hz.

The design fits the O-RAN interfaces without new signaling. Once per statistics window, a control-plane message carries the $B\times \Nr$ coefficients of $\bW$, about 12 kilobytes at $\Nr=128$, $\Bmax=24$, and 16-bit resolution, negligible against the per-slot stream payload. The antenna-level SRS samples add about 2\% to that payload for one SRS symbol per 10\,ms at 30\,kHz subcarrier spacing. In prior work, we showed that single-instruction, multiple-data (SIMD) vectorization meets the per-slot deadlines on such hardware~\cite{park2026simd}.

The first limitation of the allocation is the size of its gain over the strongest baseline, the threshold-based JSDM allocation. The greedy exceeds JSDM by less than $0.1$\,bps/Hz at $\Users=8$, a difference of no practical size, and by $1.3$\,bps/Hz at the largest configuration of \tabref{tab:scale}. Over that baseline, the allocation contributes an optimality guarantee for the capture score, the objective that the loss analysis characterizes, rather than a large rate gain, and the exact-rate check of \secref{sec:checks} shows that maximizing that score costs almost no rate. The second limitation concerns coverage. Our multi-cell evaluation places one interferer per user, assumes the interferer covariance known only for the subtraction, uses the limiting contaminated covariance in \figref{fig:multicell}, and gives the DU the exact per-slot interference-plus-noise covariance with an uncontaminated reduced channel, which requires demodulation reference signals that are not reused across cells. The third concerns the short-timescale receiver, which uses the exact reduced-dimension channel. In the 7.2x split the DU estimates that channel from the forwarded pilot streams, so the estimation error and the pilot overhead, which scale with $B$ rather than with $\Nr$, remain to be included.
\section{Conclusions}
\label{sec:conclusion}

We studied uplink combining in an XL-MIMO system whose fronthaul carries a fixed number of reduced-dimension streams. We developed a two-timescale design that allocates the per-user combining ranks from the channel statistics under a total stream budget, and applies the instantaneous MMSE receiver in the reduced dimensions. For disjoint visibility regions, we proved that the reduced-dimension receiver decouples exactly across users. From that decoupling, we derived a second-order characterization of the sum-rate loss, which matched the simulated loss on a disjoint-visibility geometry at all but the tightest budgets. We showed that the saturating-capture greedy minimizes the dominant loss term exactly. In simulations the allocation gained over the uniform baseline, matched a budget-trimmed threshold rule at eight users and exceeded it at the largest configuration. Under out-of-cell interference the gain survived with clean covariances where the interferer windows were offset by at least half a width. Window-restricted covariance estimation removed the pilot contamination where the windows were disjoint. The gain followed the heterogeneity of the per-user eigenvalue tails and vanished when the users were alike.

A tighter loss characterization for overlapping visibility regions is a topic for future work. The corrected bound~\eqref{eq:leak} was never violated in our tests but stays one to two orders of magnitude above the simulated loss. We did not derive a closed-form bound on the residual $\varepsilon$ in~\eqref{eq:char}. An analytical boundary for the regime in which the allocation gains over threshold rules is also open, since the heterogeneity sweep locates this regime only empirically. An interference-aware allocation that spends part of the budget outside the kept eigenmodes is possible, because the architecture accepts any rows in $\bW$. Temporal aging of the long-term statistics, the untested half of the two-timescale premise, remains to be studied. Near-field wavefronts, which the visibility-region model does not cover, and cell-free deployments are extensions.

\appendices
\section{Proof of Theorem~\ref{thm:decouple}}
\label{app:decouple}
\begin{IEEEproof}
We first translate the hypothesis into orthogonality. The covariances are Hermitian, so $(\bR_j\ba)^*(\bR_u\bc) =\ba^*\bR_j\bR_u\,\bc=0$ for all $\ba$ and $\bc$, which gives $\range(\bR_j)\perp\range(\bR_u)$ for $j\ne u$. The channels lie in their own ranges. So do the retained eigenvectors, because $r_u\le\mathrm{rank}(\bR_u)$. Hence $\bUrH{j}\bUr{u}=\b0$ and $\bUrH{j}\bh_u=\b0$ for $j\ne u$.

The orthogonality of the ranges gives the combining matrix orthonormal rows, $\bW\bW^*=[\bU_j^{(r_j)*}\bU_{j'}^{(r_{j'})}]_{j,j'}=\bI_B$, and confines each effective channel to a single block, $\bhr{u}=\bW\bh_u=[\b0;\,\dots;\,\bb_u;\,\dots;\,\b0]$ with $\bb_u=\bUrH{u}\bh_u$. The covariance $\bPhi=\sum_{j}p_j\bhr{j}\bhr{j}^*+\sigsq\bI_B$ is therefore block diagonal, with $u$-th diagonal block $p_u\bb_u\bb_u^*+\sigsq\bIr{u}$. Because subtracting user $u$'s own term leaves $\sigsq\bIr{u}$ in block $u$ while $\bhr{u}$ vanishes outside that block, the SINR in~\eqref{eq:rate} reduces to $p_u\,\bb_u^*(\sigsq\bIr{u})^{-1}\bb_u=\eta_u\|\bb_u\|^2$, which proves the per-realization decoupling $R_u(\bW)=\log_2(1+\eta_u\|\bb_u\|^2)$.

In the eigenbasis, $\bh_u=\bR_u^{1/2}\bg_u =\bU_u\bLambda_u^{1/2}\bU_u^*\bg_u$, and $\bUrH{u}\bU_u=[\bIr{u}\ \b0]$ selects the first $r_u$ coordinates, so $\bb_u=\bLambda_{u,[1:r_u]}^{1/2}\,[g_{u,1},\dots,g_{u,r_u}]^{\Trans}$ with $\bLambda_{u,[1:r_u]}$ the leading $r_u\times r_u$ block. Taking norms, $\|\bb_u\|^2=\sum_{i\le r_u}\lambda_{u,i}|g_{u,i}|^2=\Sigma_u$ and $\|\bh_u\|^2=\sum_{i\le \Nr}\lambda_{u,i}|g_{u,i}|^2=\Sigma_u+\Sigma_u^{\perp}$. For the full-dimension reference $\bW=\bINr$, the interference matrix $\bA=\sum_{j\ne u}p_j\bh_j\bh_j^*$ satisfies $\bA\bh_u=\b0$, so $(\bA+\sigsq\bINr)^{-1}\bh_u=\bh_u/\sigsq$ and $R_u(\bINr)=\log_2\!\left(1+\eta_u(\Sigma_u+\Sigma_u^{\perp})\right)$. Both rates hold for every realization, so substituting them into~\eqref{eq:loss} and combining the logarithms inside the expectation gives~\eqref{eq:exactloss}.
\end{IEEEproof}

\section*{Acknowledgment}
The authors used AI tools to draft text based on the authors’ formulation and results and to develop simulation code using the authors’ existing code as a starting point. The authors revised and verified all AI-assisted content and take responsibility for the ideas, analyses, and conclusions.

\bibliographystyle{IEEEtran}
\bibliography{JP_xl_mimo}

\end{document}